\documentclass{ws-jai}
\usepackage[flushleft]{threeparttable}
\begin{document}

\catchline{}{}{}{}{} % Publisher's Area please ignore

\markboth{Kuiper {\itshape et al.}}{DSN Transient Observatory}

\title{DSN Transient Observatory}

\author{T.B.H. Kuiper$^\dagger$ $^\ddagger$, R.M. Monroe$^\ddagger$, L.A. White$^\ddagger$,
C. Garcia Miro$^{\S}$,
S.M. Levin$^\ddagger$, W.A. Majid$^\ddagger$ and M. Soriano$^\ddagger$}

\address{
$^\ddagger$Jet Propulsion Laboratory, California Institute of Technology,
Pasadena, California 91109, U.S.A, kuiper@jpl.nasa.gov\\
$^{\S}$NASA Madrid Deep Space Comm. Complex, INTA/ISDEFE, Madrid, Spain
}

\maketitle

\corres{$^\dagger$Corresponding author.}

\begin{history}
\received{(to be inserted by publisher)};
\revised{(to be inserted by publisher)};
\accepted{(to be inserted by publisher)};
\end{history}

\begin{abstract}
The DSN Transient Observatory (DTO) is a signal processing facility that can
monitor up to four DSN downlink bands for astronomically interesting
signals. The monitoring is done commensally with reception of
deep space mission telemetry.  The initial signal processing is done with two 
CASPER\footnote{Collaboration for Astronomy Signal Processing and Electronics
Research} ROACH1 boards, 
each handling one or two baseband signals.  Each ROACH1 has a 10~GBe interface 
with a GPU-equipped Debian Linux workstation for
additional processing. The initial science programs include monitoring Mars for
electrostatic discharges, radio spectral lines, searches for fast radio bursts 
and pulsars and SETI. The facility will be available to the scientific community
through a peer review process.
\end{abstract}


\section{Introduction}

\noindent The NASA Deep Space Network (DSN) is a world-wide 
facility for communicating with deep space missions. Transient Observatory 
(DTO) is a signal processing facility that can monitor up to four DSN telemetry 
bands for astronomically interesting
signals. Since telemetry signals occupy only a small fraction of the entire band,
the rest of the band can be used for commensal scientific investigations.  The 
signal processing requirements may be quite different for these projects. DTO 
supports these by using different firmware on the same platform.
 
Currently, the primary scientific objective is monitoring Mars for electrostatic
discharges. Such bursts, reported by \citet{ruf09}, occurred during a convective
dust storm.  Since the DSN is receiving data from spacecraft at Mars almost
continuously, a statistically significant database can reveal the conditions
which give rise to such events.  Other planets can also be monitored for 
transient phenomena.

Additional investigations include searching for astronomical transients such as
fast radio burts (FRBs), pulsars and signals resulting from extraterrestrial
technologies.

There is one 70m diameter antenna in Goldstone, California, with others
near Canberra, Australia, and Madrid, Spain, roughly
equidistant in longitude to enable
continuous communication with robotic missions exploring the Solar System.
Each site also has at least three 34-m antennas.  Table~\ref{tab:bands}
\begin{wstable}[ht]
\caption{Deep space to earth communication bands used by the 
DSN.}
\begin{tabular}{@{}cccc@{}} \toprule
Band & Frequencies   & Bandwidth & Polarization \\
     & (GHz)         & (MHz) \\ \colrule
 S   & 2.265 - 2.305 & 40  & LCP $|$ RCP \\
 X   &  8.25 - 8.65  & 400 &  LCP \& RCP\\
 Ka  & 31.85 - 32.25 & 400 & LCP \& RCP \\ \botrule
\end{tabular}
\begin{tablenotes}
\item[a] The polarization for S-band is configured at the beginning of a 
tracking session.  It is normally RCP.
\item[b] Only some 34-m antennas have S-band.
\end{tablenotes}
\label{tab:bands}
\end{wstable}
shows the bands used for deep space to Earth communications.

The initial signal processing is done with two ROACH1\footnote{Reconfigerable 
Open Architecture Computing Hardware, model 1} boards, each 
handling up to two 640MHz-wide baseband signals.  Each ROACH1 has a 10~GB
ethernet (10~GBe) interface with a GPU-equipped Debian Linux workstation for
additional processing. The monitoring is done commensally with reception of
deep space mission signals.

Since deep space missions move slowly across the sky, the time for
acquiring data from a given direction can be quite long, though the direction is
not under the investigator's control.  At minimum elongation, Venus moves about 
one beamwidth 
per hour at X-band for a 70m antenna or at Ka-band for a 34m antenna. 
For Mars the minimum dwell time is about 100~min. For Jupiter and Saturn it is
about 5$\frac{1}{2}$ and 11~hr respectively.   If, one the other hand, one
wishes to survey more sky quickly, for a 34m antenna that is not moving,
the dwell time for a position on the sky is $\sim50$s at S-band, $\sim14$s at 
X-band and 3.8s at Ka-band.
A non-tracking 34-m antenna can cover 1\% of the sky in 17 earth rotations at 
S-band, 59 at X and 227 at Ka. 

DTO can also be used during tracking sessions assigned to observations of radio
astronomical sources. The facility will be available to the astronomical
community through a peer-reviewed proposal process.

\section{Hardware}
\begin{figure}[ht]
\begin{center}
\includegraphics[width=4.9in]{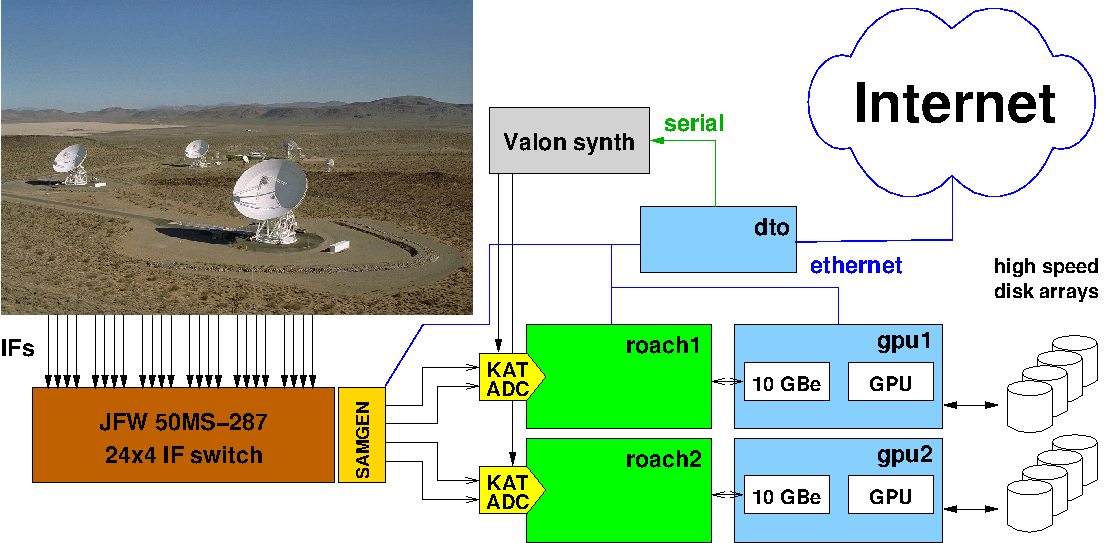}
\caption{Overview of the DSN Transient Observatory.}
\label{fig:overview}
\end{center}
\end{figure}

\noindent Figure~\ref{fig:overview} shows an overview of the signal processing
facility.
A commercial $24\times4$ intermediate frequency (IF) matrix switch made by JFW 
Industries, Inc., directs up to four station IFs to the inputs of
two ROACH1\index{ROACH} signal acquisition and processing units.  Two high-speed, 
multi-core, GPU\footnote{graphics processing units}-equipped workstations 
accept processed data from each ROACH over a 10~Gb ethernet (10Gbe) 
port.

The PowerPCs\index{PowerPC} on the ROACH boards boot their kernel images from 
the master controller {\ttfamily dto}. {\ttfamily dto} also has the file systems
of the PowerPCs which control the operation of a Xilinx XC5VSX95T
field programmable gate array (FPGA).
The bit-files which configure the ROACH FPGAs are loaded by the 
PowerPCs. So, in effect, {\ttfamily dto} provides the firmware for the FPGAs,
allowing rapid switching of firmware under computer control.

\begin{figure}[ht]
\begin{center}
\includegraphics[width=4.9in]{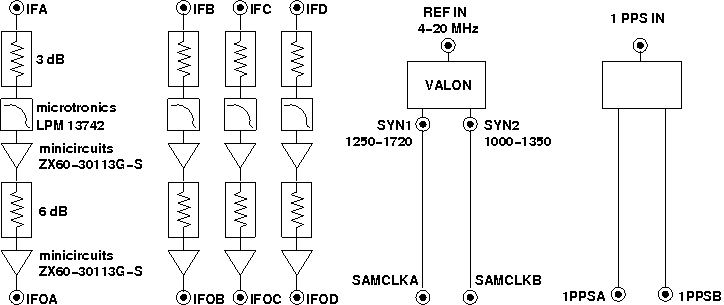}
\caption{Schematic diagram of the sampler clock and IF distribution chassis.}
\label{fig:samgen}
\end{center}
\end{figure}

The Sample Clock Generator (SamGen) subsystem provides the clock
signals for the analog-to-digital convertors (ADC), optional synchronization 
pulses and IF
amplification and low-pass filtering.  Figure~\ref{fig:samgen} has a schematic
diagram.
SamGen has a Valon dual synthesizer so
the two ROACH board ADCs can be independently clocked.  It also distributes 1pps
synchronization pulses to the KAT ADCs.

In addition to the controller {\tt dto} and the two PPCs, there are two
high-performance computers. Each post-processing PPC has two Intel Westmere 
2.66GHz X5650 processors
with six cores, allowing it to run twelve threads. It has 96 GB of RAM,
two GTX 580 GPU processors each with 512 cores, and 48TB of fast disk. 32T are
available for data storage.  All hosts have the Debian Linux operating system.

\section{Firmware}

\subsection{Kurtosis Spectrometer}

\begin{figure}[h!tb]
\begin{center}
\includegraphics[width=7in]{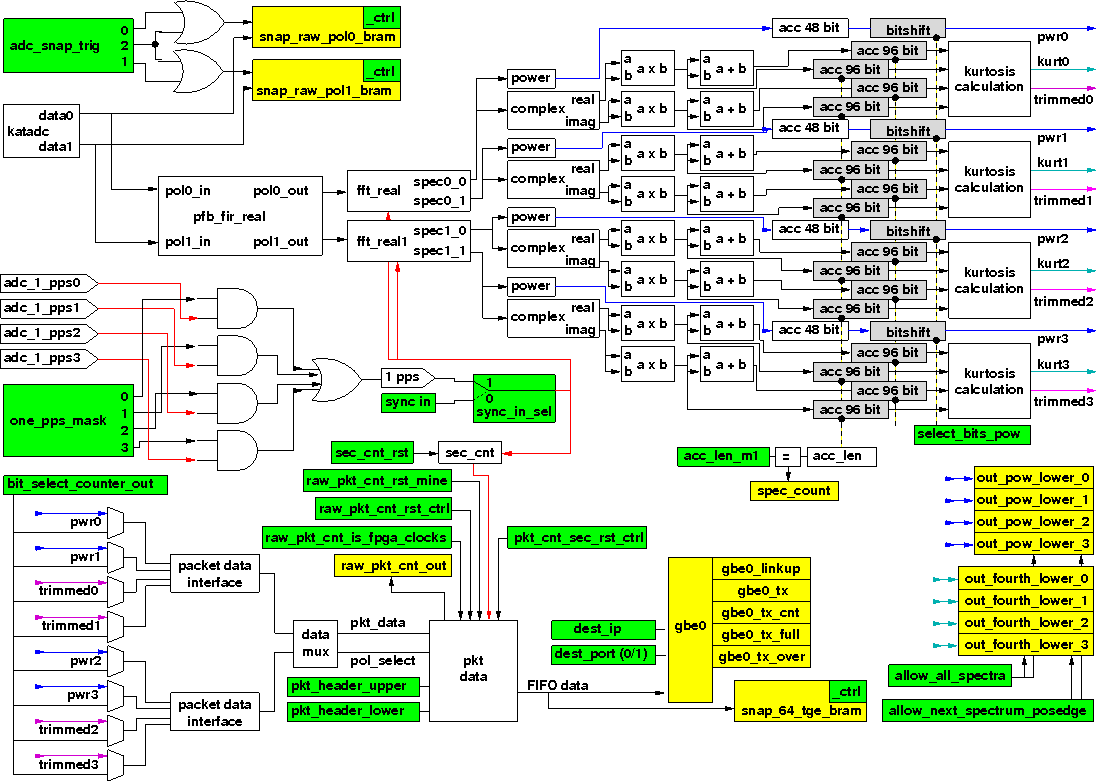}
\caption{\label{fig:kurtspec}Schematic diagram of the kurtosis spectrometer
firmware, including all the registers and data stores.}
\end{center}
\end{figure}

\noindent On 2006 June 8 Deep Space Station 13 (DSS-13) detected non-thermal 
bursty emissions from Mars \cite{ruf09} that showed resonances similar to 
Schumann resonances on Earth \cite{schumann52}.
Transient non-thermal radio bursts can be distinguished from Gaussian thermal
noise by computing the kurtosis (normalized fourth moment) of the signal
voltage \cite{ruf2006}. For such studies we have designed specifically for DTO 
a 1024-ch firmware spectrometer that computes kurtosis as well as power.  The 
design is shown in Figure~\ref{fig:kurtspec}. 
For Mars, the integration time for these calculations, a parameter which can be
adjusted, is 1~ms.  This allows a signal kurtosis spectrogram to be examined 
for the presence of modulation by very low frequency waves propagating
between the surface and ionosphere of Mars.
The DSN has telemetry with at least one spacecraft at Mars for an average of
20~hrs every day.  One of DTO's main goals is to monitor Mars diligently for
further evidence for electrostatic discharges.

\subsection{High Resolution Spectrometer}

\noindent It would be unlikely to detect radio spectral line emission from an 
arbitrary direction in the sky but planetary atmospheres can produce such 
emission from very low density upper atmospheres \cite{{gulkis1978,wilson1981}}.
One of the available firmwares is a 32K-channel spectrometer. It was
developed for the Tidbinbilla AGN Maser Survey (TAMS) project led by Harvard
Smithsonian Astrophysical Observatory \cite{2014atnf.prop.6463Z} and is part of 
the 17-27~GHz receiver system on DSS-43 \cite{kuiper2016}. With a 
640~MHz bandwidth this gives a spectral resolution of 20~kHz.  The Doppler
velocity resolution is 2.5~km~s$^{-1}$ at S-band, 0.7 at X-band and
0.2~km~s$^{-1}$ at Ka-band. This firmware is most likely to be used for
antenna time assigned to astronomical research. It would also be useful for
monitoring radio frequency interference (RFI) and spacecraft transmitters which
have lost lock and drifted in frequency.

\subsection{SETI Spectrometer}

\noindent Extraterrestrial civilizations that are located close to the Sun's
ecliptic plane could be aware of Earth and its life-supporting
potential from Earth's transits across the Sun.  Such a civilization might
direct a beacon towards Earth to initiate possible contact.  An interesting
strategy is to conduct searches for extraterrestrial civilizations (SETI) along
the ecliptic plane \cite{henry2008}.  The DSN, in communication with 
spacecraft at or travelling
towards other planets, has many antennas pointed close to the ecliptic.
A commensal SETI search may have enhanced chances of success compared to
similar searches at radio observatories \cite{werthimer2001}. There are 82 
K and G stars within
one kpc of Earth within this zone \cite{2016AsBio-16-259H}, so that over a time 
comparable to the orbital periods
of the planets, they would be within the beam of antennas communicating with
deep space missions.

Amplitude modulated signals transmitted by putative extraterrestrial 
civilizations are expected to be most easily detected by the carrier tone, which
is expected to be extremely narrow.  An extremely high resolution spectrometer
design, 
SERENDIP~VI\footnote{//casper.berkeley.edu/wiki/SETI/FRB\_Spectrometer\_(SERENDIP\_VI)},
will be adapted from ROACH2. The firmware performs 4K channelization.  The
samples from each channel of the firmware output are further 
channelized in software using GPUs. A 128K PFB
on each channel from the firmware spectrometer will yield a spectral resolution
of 1.2~Hz. SERENDIP~VI has been implemented in SETIBURST, a multi-purpose
signal processor \cite{chenna2016} similar in concept to DTO.

The high resolution of the final spectra allows masking of spacecraft telemetry
and known interference signals.  Since the amount of data is still huge, 
additional back-end processing will decimate the data. For example, the JPL 
``setispec'' used in the GAVRT SETI 
project\footnote{http://galileo.gavrt.org/seti/} \cite{jones2010}, reports for 
each of 4096 coarse channels, the number of and power detected in the strongest
high-resolution sub-channel. 

\section{Software}

\subsection{Monitor and Control}

\noindent The monitor and control (M\&C) software has a server/client 
architecture.
All control software is written in Python.
Inter-process communication (IPC) is managed with 
Pyro\footnote{http://pythonhosted.org/Pyro4/}.

\begin{figure}[h!tb]
\begin{center}
\includegraphics[width=4.9in]{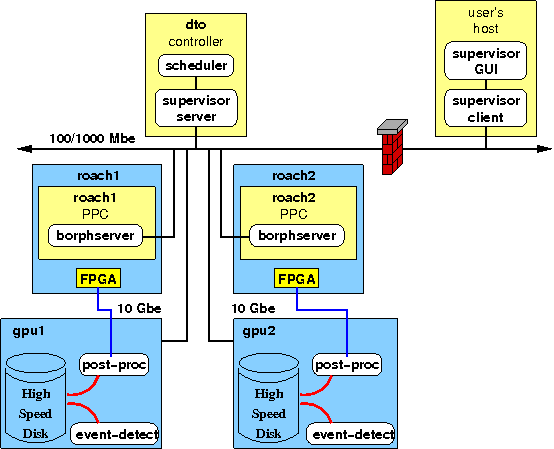}
\caption{\label{fig:sw_oview}Overview of the DTO software.}
\end{center}
\end{figure}
Figure~\ref{fig:sw_oview} gives a schematic overview of the software.
The {\bfseries scheduler} processes the DSN 7-Day Schedule 
for the Goldstone antennas
and selects antennas and frequency bands according to a rule set determined
by the science team.

The {\bfseries supervisor} is a client of the hardware servers and
monitors and controls all aspects of DTO --
the state of the IF switches, the frequency and power of the sample clock
generators, the gain of the KATADC RF sections, the firmware loaded into
the FPGAs, etc.  It also collects metadata from the station's monitor data
server using the DSN's IPC protocol.

The can be overseen by a human through the supervisor 
client.
This has a set of classes and functions through which an expert user can
communicate with the supervisor through a Python command line.  
A graphical
user interface (GUI) can invoke the same classes and functions to provide
a more convenient overview.

The {\bfseries ROACH Power PCs} control most of the ROACH operation.
The PPC filesystems are on {\ttfamily dto}.   The supervisor loads the 
appropriate firmware for a session into the ROACHs.

The user's client software works through a tunnel \cite{kuiper2012}
which is created by a user with
the requisite authority (username and two-factor passcode) to log in to the 
firewall's gateway.  Only one log-in is needed for a session.

\subsection{Post-Processing}

Data generated by the firmware streams through 10GbE connections to the signal 
processing hosts.  Because of the generally high data rate
involved, this software is normally written multi-threaded C. Python's libraries
for GPU support are being investigated. The ouput
is written to disk for further real-time analysis.

In the final software, the data are stored temporarily in a circular buffer.
These data are examined for ``events''.
This software is the most complex as it must separate real from false
events, generate alerts for events requiring prompt attention and store
data which  events of interest.

\section{Status and Conclusion}

\noindent DTO is awaiting shipment to the Goldstone Deep Space Communications
Complex (GDSCC).  The hardware, firmware and
M\&C software for the initial science program has been verified in the laboratory at
JPL. The location and access to power and signals at Goldstone have been assigned.
Paperwork required for installation in the configuration-controlled DSN 
operations environment has been completed.
DTO will become operational at Goldstone in the late summer of 2016.

DTO will be operated as an open facility.  The JPL Interplanetary Network
Directorate (IND) which manages the DSN will accept proposals from outside
investigators who bring their own firmware to conduct research with different 
goals.

\section{Future Work}

Because of the flexibility built into the CASPER development ecosystem, DTO 
should not be considered limited to the applications described here, but rather 
a prototype for general purpose commensal science signal processing. For 
example, the high resolution spectrometer design was recently implemented as 
four spectrometers on a single ROACH2. Additional modifications, such as 
extending the kurtosis based designs for flexible integration times, and 
development of higher order (6th, 8th moment) statistics are also ongoing. 

The Instituto Nacional de T\'{e}cnica Aeroespacial (INTA) and Ingeniería de Sistemas 
para la Defensa de Espa\~{n}a (ISDEFE) have made available four ROACH1 boards together
with a controller host to the Radio Astronomy Department of the Madrid
Deep Space Communications Complex (MDSCC). INTA and JPL will collaborate
in assembling a second DTO for MDSCC which will be operated by INTA scientists.

A fifth ROACH1 board will be used to provide a spectrometer for the PARTNeR 
educational antenna (Proyecto Acad\'{e}mico del RadioTelescopio de NASA en Robledo), 
a project also managed by INTA.

\section*{Acknowledgments}
\noindent The Deep Space Network is operated by the California
Institute of Technology Jet Propulsion Laboratory for the National
Aeronautics and Space Administration (NASA).
This work was funded by the Science and Technology Investment program 
of the offices of the JPL Chief Scientist and Chief Technologist.

We are endebted to Jonathan Kocz for completing the kurtosis firmware's
10~Gbe output capability.
We thank Chris Ruf and Nilton Renno for insightful discussions on kurtosis and
the science of electrostatic discharges in the Martian atmosphere. The advice 
and support of the JPL Interplanetary Network Directorate Chief Scientist,
Joseph Lazio, is gratefully acknowledged.
Chuck Naudet leads the DSN Science Automation task which develops software
that will enhance DTO operation. Dong Shin assisted in assembling DTO.


\begin{thebibliography}{9}

\bibitem[Chennamangalam et al.(2016)]{chenna2016} Chennamangalam, J.,
MacMahon, D., Cobb, J., et al.\ 2016, submitted to {\it ApJS}.

\bibitem[Gulkis et al.(1978)]{gulkis1978} Gulkis, S., Janssen, M.~A., \& 
Olsen, E.~T.\ 1978, {\it Icarus}, {\bf 34}, 10. 

\bibitem[Heller \& Pudritz(2016)]{2016AsBio-16-259H} Heller, R., \& Pudritz, 
R.~E.\ 2016, {\it Astrobiology}, {\bf 16}, 259 

\bibitem[Henry et al.(2008)]{henry2008} Conn Henry, R., 
Kilston, S., \& Shostak, S.\ 2008, {\it  BAAS}, {\bf 40}, 4.06.

\bibitem[Jones et al.(2010)]{jones2010} Jones, G., Weinreb, S., Mani, H., 
et~al.\ 2010, {\it ProcSPIE}, {\bf 7733}, 77333O.

\bibitem[Kuiper {\it et al.}(2012)]{kuiper2012} Kuiper, T. B. H., Majid, W. A.,
Martinez, S., Garcia-Miro, C., Rizzo, J. R. [2012], in
Observatory Operations: Strategies, Processes, and Systems IV 
{\it Proceedings of the SPIE}, {\bf 8448}.

\bibitem[Kuiper {\it et al.}(2016)]{kuiper2016} Kuiper, T.B.H., Franco, M.,
Smith, S., Baines, G., Greenhill, L., Horiuchi, S., Olin, T., Price, D.,
Soni, A., Teitelbaum, L.P., Weinreb, S., and White, L., Zaw, I., 2016,
{\it JAI}, in preparation.

\bibitem[Ruf {\it et al.}(2006)]{ruf2006} Ruf, C. S., Gross, S. M., Misra, S.,
IEEE Trans. Geosci. Remote Sens., {\bf 44}, 694.

\bibitem[Ruf {\it et al.}(2009)]{ruf09} Ruf, C., Renno, N. O., Kok, J. F., 
Bandelier, E., Sander, M. J., Gross, S., Skjerve, L. and Cantor, B. [2009], 
{\itshape Geophys. Res. Lett.}, {\bf 36}, L13202.

\bibitem[Schumann(1952)]{schumann52} Schumann, W. O. [1952] {\itshape Z. Nat.},
{\bf 72}, 250.

\bibitem[Werthimer et al.(2001)]{werthimer2001} Werthimer, D., Anderson, D., 
Bowyer, C.~S., et al.\ 2001, {\it ProcSPIE}, {\bf 4273}, 104.

\bibitem[Wilson et al.(1981)]{wilson1981} Wilson, W.~J., Klein, M.~J., 
Kakar, R.~K., et al.\ 1981, {\it Icarus}, {\bf 45}, 624 

\bibitem[Zaw et al.(2014)]{2014atnf.prop.6463Z} Zaw, I., Greenhill, L., 
Briggs, F., Moin, A., Horiuchi, S., Kuiper, T., and Soni, A. 2014, 
ATNF Proposal 6463Z.


\end{thebibliography}
\end{document}